\documentclass[conference]{IEEEtran}
\IEEEoverridecommandlockouts
\usepackage{cite}
\usepackage{amsmath,amssymb,amsfonts}
\usepackage{graphicx}
\usepackage{textcomp}
\usepackage{xcolor}
\usepackage{enumitem}
\usepackage{algorithm}
\usepackage{algpseudocode}

\usepackage{qcircuit}
\usepackage{physics}
\usepackage{subcaption}

\def\BibTeX{{\rm B\kern-.05em{\sc i\kern-.025em b}\kern-.08em
    T\kern-.1667em\lower.7ex\hbox{E}\kern-.125emX}}
\begin{document}

\title{A Quantum Computing Approach for Multi-robot Coverage Path Planning}

\author{\IEEEauthorblockN{ Poojith U Rao, Florian Speelman, Balwinder Sodhi, Sachin Kinge}}


\maketitle

\begin{abstract}


This paper tackles the multi-vehicle Coverage Path Planning (CPP) problem, crucial for applications like search and rescue or environmental monitoring. Due to its NP-hard nature, finding optimal solutions becomes infeasible with larger problem sizes. This motivates the development of heuristic approaches that enhance efficiency even marginally.

We propose a novel approach for exploring paths in a 2D grid, specifically designed for easy integration with the Quantum Alternating Operator Ansatz (QAOA), a powerful quantum heuristic. Our contribution includes:

\begin{itemize}
    \item An \textit{objective function} tailored to solve the multi-vehicle CPP using QAOA.
    \item \textit{Theoretical proofs} guaranteeing the validity of the proposed approach.
    \item Efficient \textit{construction of QAOA operators} for practical implementation.
    \item \textit{Resource estimation} to assess the feasibility of QAOA execution on current hardware.
    \item \textit{Performance comparison} against established algorithms like the Depth First Search, demonstrating significant runtime improvements.
\end{itemize}

This work paves the way for leveraging quantum computing in optimizing multi-vehicle path planning, potentially leading to real-world advancements in various applications.

\end{abstract}


\section{Introduction}

Advancements in robotics, drones, sensors, and navigation technology have led to the widespread adoption of multi-vehicle systems for numerous applications.
The common objectives include surveillance, monitoring, surveying, and modeling in landmine detection, lawn mowing, search and rescue operations, and natural disaster detection. Such objectives can be achieved more efficiently when multiple vehicles communicate and cooperate. The common aim of such applications is to scan an area of interest while optimizing various objective functions. Finding an optimal and collision-free route to cover an area of interest is the Coverage Path Planning (CPP) problem.

The two main components of CPP include i) viewpoint generation and ii) path generation \cite{almadhoun2019survey}. \textit{Viewpoint generation} refers to identifying important positions in the area to be covered. The viewpoints help in collecting data that aid in covering the entire area. Most of the algorithms in the literature take a uniform distribution of viewpoints across the Area of Interest (AoI). A simple example is the 2D grid-based approach, where viewpoints are modeled as a grid with different shapes, such as triangles, squares, hexagons, etc. A more detailed study on different viewpoint generation approaches is discussed in \cite{almadhoun2019survey}.

\textit{Path generation} refers to finding an optimal route that covers these viewpoints. This step starts with dividing the AoI into sub-areas and initializing the robot's location. Then, the covering direction is determined for each sub-area. When this is known, each sub-area is covered using simple movements. The efficacy of the path identified can be measured in terms of numerous parameters such as the percentage of covered area, path overlap rate, and energy consumption of robots \cite{tan2021comprehensive}.

CPP can be viewed as a combinatorial optimization problem with multiple objectives and constraints depending on its application. Commonly used objective functions include the number of rotations or turns in the path, number of viewpoints covered, computing time, path length, path overlap, energy consumption, smooth trajectories/maneuvers, etc., and constraints include avoiding obstacles, no backtracking, a continuous path, usage of simple trajectories such as straight lines and circles, etc.


Developing new algorithms for solving combinatorial optimization problems with hard constraints, such as CPP, that can improve the runtime performance even by a small fraction is of great importance. It is because, at scale, even a slight improvement in resource optimization accrues significantly in a large-scale real-world application scenario.

\subsection{Related work}

Multiple works aim to solve the Coverage path planning using a single robot. In the case of a large area of interest (AoI), single robot solutions are not preferred due to the high possibility of malfunction, mechanical or electronic breakdown, sensor and actuator faults, and battery drainage. Thus, recent research uses multiple robots to overcome most of these shortcomings \cite{sun2019multi}. 
Numerous challenges associated with the CPP problem along with an exhaustive, comprehensive review of existing algorithms used for solving it exist in literature \cite{tan2021comprehensive}. Some of the relevant works where a multi-vehicle version of the CPP problem are discussed here.

Spanning trees have been used in numerous works for solving the multi-vehicle version of CPP \cite{senthilkumar2008spanning, agmon2006constructing,gao2019stc}. They can be used as a base for creating coverage paths and a polynomial-time tree construction algorithm is used to dramatically improves the coverage time \cite{agmon2006constructing}. However, the traveled path in this algorithm depends on the initial positions of each robot, which might lead to issues such as backtracking among other robots, a high overlap rate, significantly smaller energy efficiency.

Artificial Potential Fiend (APF) method has also been used to cover the area and find collision-free paths for the multi-robot system \cite{huang2018potential}. Despite all the research efforts, there is still a lack of planning for collision avoidance between multiple robots when simultaneously accessing the goal under the potential field.

Dijkstra’s algorithm is also proposed for task planning in a multi-robot system. Nevertheless, the search path is not optimal in terms of travel distance \cite{rosa2020using , zhang2019optimal}. Theta $A^*$ algorithm was used in \cite{huang2018development} to reduce the coverage time. However, the algorithm failed to generate a global optimization solution in terms of path length. The Voronoi partition approach is another common modeling technique applied in the distributed coordination for a multi-robot system \cite{nair2020gm}.


Genetic Algorithm \cite{sun2019multi}, Bee Colony Optimization (BCO) \cite{caliskanelli2014multi}, Firefly algorithm (FA) \cite{de2015bio , palmieri2015multi}, Coordinated Multi-robot Exploration (CME), Gray Wolf Optimization (GWO) algorithm \cite{albina2019hybrid} and numerous other evolutionary algorithm based solutions have been proposed for the CPP. The performance of FA, PSO, and BCO in coordinating the swarm robotics system in terms of energy consumption was compared by \cite{palmieri2019comparison}. The prevalence of heuristic approaches for the CPP problem underscores their real-world practicality and value. Hence, in this work, heuristic based approach capable of taking advantage of quantum mechanics is proposed.

\label{sec:modeling-cpp}

The multi-vehicle version of the Coverage Path Planning problem is NP-hard. Thus, mostly heuristic-based \cite{stollenwerk2020toward} solutions are proposed that can neither guarantee optimality nor shorten running times. There is a recent trend \cite{stollenwerk2020toward} in using quantum algorithms on near-term quantum devices that promise potential quantum advantage for solving combinatorial problems. While more attention is given to quantum annealing due to the near-term availability of hardware, recent works also show the applicability of gate-based quantum computers for solving such problems. In particular, for solving problems with hard constraints on gate-based computers, the quantum alternating operator ansatz (QAOA) algorithm was proposed by \cite{hadfield2019quantum}.

In this work, an approach to explore paths between two points in a 2D grid is proposed. The approach involves starting from a trivial initial path (random state) and exploring nearby solutions until a satisfactory solution is found within the feasible solution space. The methodology proposed to explore new solutions can be adapted to other well-known heuristic algorithms. For example, it can be implemented as mutations/crossover in Genetic Algorithms (GA), random neighbors in Simulated Annealing (SA), and mixers in Quantum Alternating Operator Ansatz (QAOA).

In this work, the focus is on the Quantum Alternating Operator Ansatz (QAOA), which is a meta-heuristic framework proposed by \cite{hadfield2019quantum} used to perform approximate optimization on gate-based quantum computers. This is an extended version of the Quantum Approximation Optimization Algorithm proposed by \cite{farhi2014quantum} that produces approximate solutions for combinatorial optimization problems. The version proposed in \cite{hadfield2019quantum} is well suited for problems such as CPP, where the feasible space is smaller than the complete solution space. The QAOA algorithm can be summarized as follows:

\begin{enumerate}
    \item Set the quantum system to an \textit{initial state}, which is trivial to implement and lies inside the feasible subspace of the problem.
    \item Apply suitable parameterized quantum operators $p$ times on the initial state.
    \item Measure the system in the computational basis to find a candidate solution to the problem.
    \item Update the parameters of the operators to optimize the cost function.
    \item Repeat 2-3-4 until desired convergence is obtained.
    \item Measure in the computational basis to find a solution to the problem.
\end{enumerate}

In the following sections, the methodology to explore paths in a 2D grid is described. The methodology is described in terms of a Genetic Algorithm. Later, the design of mixer operators and phase separation operators for the QAOA algorithm, along with approximate resource estimates, are discussed. The proposed methodology is evaluated by implementing the Simulated Annealing and QAOA algorithm and comparing it with the DFS algorithm.

\section{Coverage Path Planning problem}
\label{sec:cpp-description}

In this section, an approach based on Genetic Algorithms (GAs) is proposed for exploring all paths between two nodes in a 2D grid. As with any GA, the approach starts with an initial population and explores new feasible paths by selecting individuals from the current population to be parents and produce children for the next generation. An objective function evaluates the quality of the population. Over successive generations, the population evolves towards an optimal solution. The important steps of the algorithm are described in the following subsections.

\subsection{Problem definition}

The Area of Interest (AOI) is defined by a rectangular grid of dimensions $n \times m$ and characterized by a set $E$ of edges. There are $r$ robots, denoted as $r_i$, each initially stationed at $s_{r_i}$ (source station). This can be visualized as a cube-shaped graph of size $r \times n \times m$. The robots navigate the AOI along the grid edges from the source node to the destination node. Every node within the grid serves as a viewpoint and is to be covered by the robots. The path of a robot is represented by a sequence of edges from the source node to the destination node.

The objective of the multi-vehicle CPP is to allocate paths to the robots with the following criteria: i) maximize coverage of viewpoints, ii) minimize the total time required by all robots for coverage, and iii) minimize redundant visits to a node by different robots.

The paths assigned to each robot must also adhere to the following conditions: i) avoid backtracking while exploring viewpoints, ii) be continuous and sequential, and iii) navigate around obstacles on the map.

\subsubsection{Decision variable}

Robots move along the edges of the graph to cover all viewpoints. Each edge in the grid is considered as a decision variable (Equation \ref{eq:cpp-dec-var}). The variable represents if an edge is covered by a robot. In a grid of size $n \times m$, there are $m (n - 1) + n (m-1)$ edges. Hence, the total number of variables needed is $r(m (n - 1) + n (m-1))$.

\begin{equation}
    \label{eq:cpp-dec-var}
    x_{i,e} = 
    \begin{cases}
        1 & \text{if the robot i covers edge e}  \\
        0 & \text{otherwise} \\
        & i = 1,\cdot \cdot \cdot, r ; e \in E; \\
    \end{cases}
\end{equation}

\subsubsection{Objective function}

The objective function has to satisfy four requirements: 
\textbf{R1)} avoids obstacles along the path, \textbf{R2)} maximizes the number of viewpoints covered, \textbf{R3)} minimizes the overall cover time needed and \textbf{R4)} minimize overlapping of paths assigned to robots.

Obstacles along the path can be avoided by avoiding all the edges leading towards it. To satisfy \textbf{R1)} and \textbf{R2)}, a weight is assigned to each edge in the graph. A positive weight is assigned to all edges leading to obstacles, and a negative weight is assigned to the remaining edges. When the objective function is minimized, the positively weighted edges are preferred over negatively weighted edges. The cost function for \textbf{R1)} and \textbf{R2)} is given in Equation \ref{eq:cpp-cost-1} where $w_i$ is the weight assigned to $i^{th}$ edge.

\begin{equation}
\label{eq:cpp-cost-1}
    c_1 = \sum_{r,i} w_i x_{r,i}  
\end{equation}

The total time the robots take to travel the assigned paths is the cover time. An optimal time coverage algorithm for a system with $r$ robots and $m \times n$ viewpoints will result in a total coverage time of  $\frac{m \times n}{r}$ \cite{agmon2006constructing}.

When robots have equal speed, they cover equal distances in equal time. Hence, under the assumption of equal speed, it is sufficient to minimize the difference in distance covered between all pairs of robots. The cost function to compute the sum of all differences is given in Equation \ref{eq:cpp-cost-2} where $d_j$ is the length of edge $j$.

\begin{equation}
\label{eq:cpp-cost-2}
    c_2 = \sum_{r} \left [   \sum_{j} d_jx_{r,j} - \sum_{j} d_jx_{r+1,j}  \right] ^ 2
\end{equation}

Two edges are required to cover a node: one incoming edge and one outgoing edge. Hence, the sum of all decision variables representing edges connected to a node (which includes all robots) must be two (apart from source and destination nodes). This is formulated as a cost function in Equation \ref{eq:cpp-cost-3} where $j$ denotes all edges connected to node $i$.

\begin{equation}
\label{eq:cpp-cost-3}
    c_3 = \sum_{i \in \text{node}} \left [ \sum_{j \in \text{i}} \sum_r x_{r,j} -2\right ]^ 2 
\end{equation}

The final objective function is the weighted sum of all the cost functions. This is given in Equation \ref{eq:cpp-cost}.

\begin{equation}
\label{eq:cpp-cost}
    c = \alpha_0 c_1 + \alpha_1 c_2  + \alpha_2 c_3
\end{equation}

\subsection{Initial population}
\label{sec:initial-population}
The algorithm starts with an initial population consisting of paths for each robot from the source node to the destination node. A heuristic to generate the initial path is to start from the source node and move horizontally to the column of the destination node and then vertically to the row of the destination node.

\begin{figure}
    \centering
    \includegraphics[scale = 0.2]{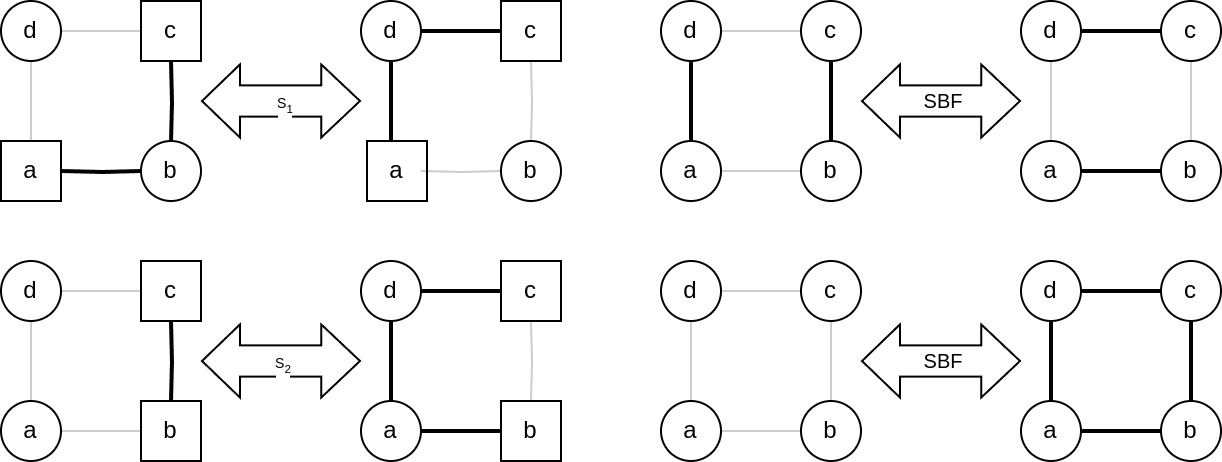}
    \caption{On the left - Scenarios where SBF operation searches for paths between two points. Top-left is the $S_1$ operation, and bottom-left is the $S_2$ operation.  On the right - Scenarios where SBF operation must be avoided}
    \label{fig:coverage-2x2}
\end{figure}

\begin{figure*}
    \centering
    \includegraphics[scale = 0.3]{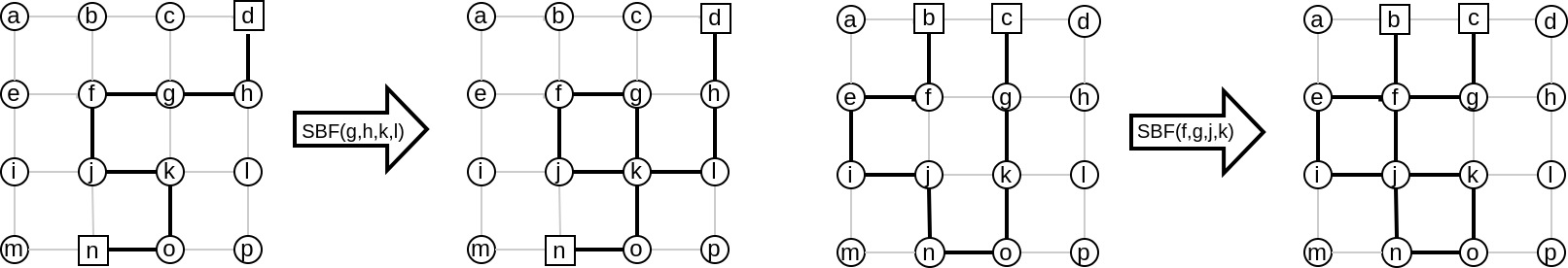}
    \caption{Scenario where SBF operation on a four-node sub-grid results in a loop}
    \label{fig:coverage-loops}
\end{figure*}

\subsection{Mutation}

The mutation operators are designed to explore all feasible solutions for the problem. A feasible solution is one where the path assigned to each robot meets the following conditions: i) It connects the source and destination nodes. ii) There is no backtracking involved. iii) An edge/node is covered at most once. iv) No loops are present. Mutations are performed only on a four-node sub-grid at a time. The following subsections show how mutations explore paths in a four-node grid and how repeated mutations on different four-node sub-grids explore all paths in the grid.

\subsubsection{Four-node graph}
A simple four-node graph is shown in Figure \ref{fig:coverage-2x2}. It has four edges $ab$, $bc$, $cd$, and $ad$ denoted by decision variables $ x_{ab}, {x_{bc}}, {x_{cd}} \text{ and } {x_{da}}$. There are exactly two paths between any two nodes in this graph. The two paths between two nodes are complements of each other. For example, if $x_{ab}{x_{bc}}{x_{cd}}{x_{da}}$ is a valid path from $a$ to $b$ then $\overline{x_{ab}} \ \overline{x_{ab}} \  \overline{{x_{bc}}} \  \overline{{x_{cd}}} \ \overline{{x_{da}}}$ is also a valid path from $a$ to $b$. Hence, if a bit string $x_{ab}{x_{bc}}{x_{cd}}{x_{da}}$ denotes a feasible path between two nodes in a four-node graph, the other feasible path is obtained by performing a NOT operation on the bit string. This operation of simultaneously flipping all the edges in a four-node graph is termed as \textbf{Simultaneous Bit Flip (SBF)} operation.

\subsubsection{Scaling to larger grids}
All feasible paths between two nodes in an $m \times n$ grid can be explored by applying SBF operation on smaller $(m-1)(n-1)$ four-node sub-grids. On applying an SBF operation on a four-node grid, while the path inside the four-node grid remains valid, the path outside the four-node grid can become invalid (a discontinuous or a loopy path). Hence, specific conditions must be satisfied before an SBF operation can be applied to a four-node sub-grid. Examples of valid and invalid scenarios are shown in Figure~\ref{fig:coverage-2x2}. In the subsequent sections, the logic function to identify valid scenarios is discussed.


\subsubsection{Logic function representing valid/invalid paths inside four node sub-grid}
The logic function to identify valid scenarios inside a four-node sub-grid depends on the following scenarios:

\textit{1. There is no path in the sub-grid:}
Applying an SBF operation on a four-node grid with no active edges (Figure \ref{fig:coverage-2x2} - right bottom) forms a loop. Hence, if $x_{abcd} = {0000}$, the operation must not be applied. The logic expression for the same is $f_1(a,b,c,d) = x_{ab} \lor x_{bc} \lor x_{cd} \lor x_{ad}$.
    
\textit{2. There are two paths in the sub-grid:}
Applying an SBF operation on a four-node grid with two parallel paths (Figure \ref{fig:coverage-2x2} - right top) alters the source and destination of the paths. Hence if ${x_{abcd}} = {0101} \lor {x_{abcd}} = {1010}$ the operation must not be applied. The logic expression for the same is $f_2(a,b,c,d) = \overline{x_{ab}\overline{x_{bc}}x_{cd}\overline{x_{ad}} \lor \overline{x_{ab}} x_{bc} \overline{x_{cd}} x_{ad}}$.
    
\textit{3. There is a single path in the sub-grid:} Applying the SBF operation on a four-node sub-grid in this scenario may create loops in its adjacent sub-grids. Hence, additional cases need to be verified. If the path passes through any of the nodes in the four-node grid but does not enter it, then applying the SBF operation on the four-node grid crisscrosses two paths. An example covering this scenario is shown in Figure \ref{fig:coverage-loops}. Let $oe$ denote the edges of the nodes in the sub-grid but not in the sub-grid. If any of the $oe$ edges are active, the operation must not be applied on the sub-grid. The logical expression for the same is $f_3(a,b,c,d) = \overline{x_{a,oe_1}x_{a,oe_1} \lor x_{b,oe_3}x_{b,oe_4} \lor x_{c,oe_5}x_{c,oe_6} \lor x_{d,oe_7}x_{d,oe_8}}$.

The final logic function that decides if the operation can be applied on the four-node sub-grid is the \textit{binary and} operation of all the logical expressions. This is given as

\begin{equation}
    \label{eq:cpp-logic-function}
    f(a,b,c,d) = f_1(a,b,c,d) \land f_2(a,b,c,d) \land f_3(a,b,c,d)
\end{equation}

\begin{figure*}
    \centering
    \includegraphics[scale = 0.25]{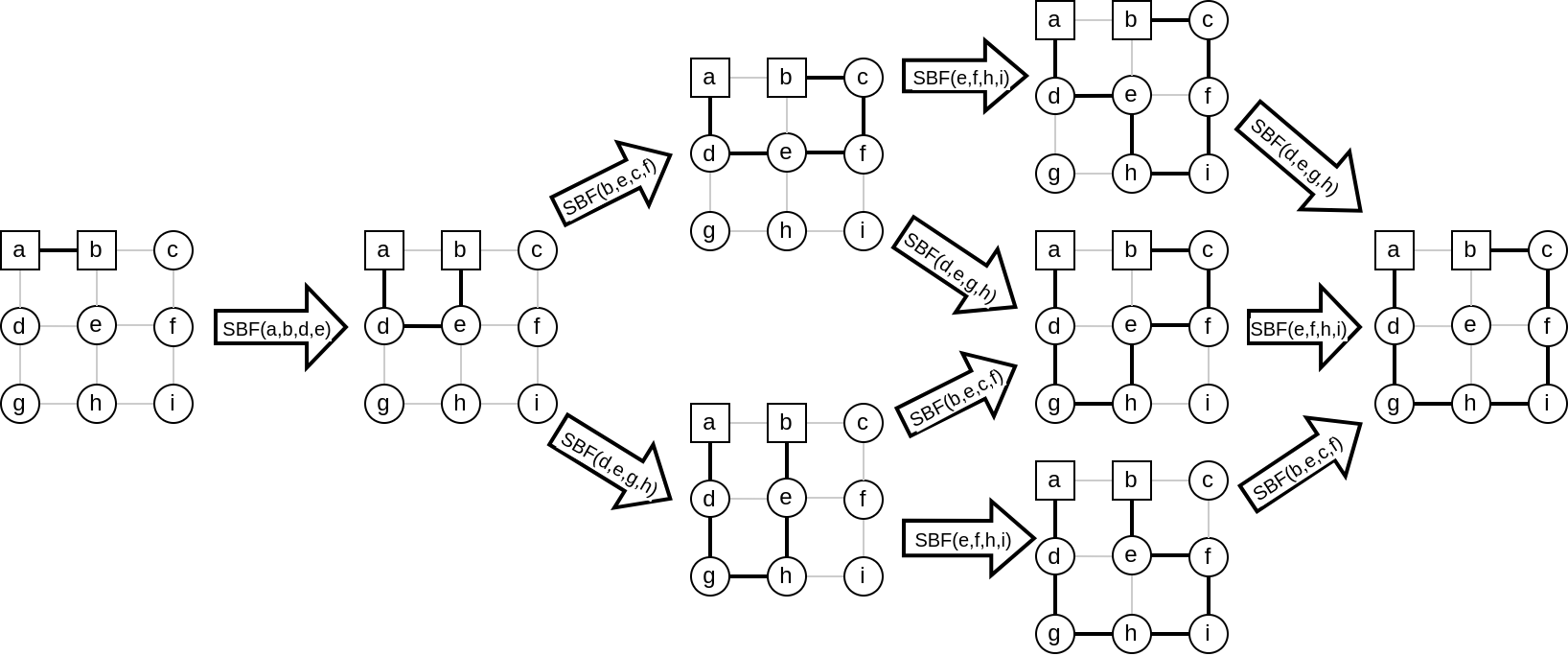}
    \caption{Exploring all feasible paths between two points $a$ and $b$ by applying SBF operation on 4-node sub-grids.}
    \label{fig:coverage-3x3}
\end{figure*}

The order of mutations controls the exploration of paths. As the population evolves with more mutations, more feasible solutions are explored. The proof that SBF operations explore all feasible solutions on four-node subgrids is discussed in the next section. An example exploration in a 3x3 grid is shown in Figure \ref{fig:coverage-3x3}.

The pseudo-code of the GA explained above is as follows:

\begin{enumerate}
    \item \textbf{Initial population:} Start with bit strings that assign a valid path to each robot.

    \item Compute the quality of the population using the objective function.
        
    \item Until a desired quality is achieved, repeat the following: 
    
    \begin{enumerate}

        \item \textbf{Mutation:} Apply the SBF operation on all four-node sub-grids and add the offspring to the population. 
        
        \item Evaluate the quality of the population using the objective function.
        
    \end{enumerate}
    
\end{enumerate}

The above methodology can be adopted into the Simulated Annealing (SA) algorithm by using the SBF operation to search for nearby neighbors. Similarly, for QAOA, the SBF operation can be incorporated into the mixer design, and the objective function can be incorporated into the phase operators. The design and construction of these operators are discussed in the subsequent sections.

\section{Proof of explorability}

\begin{figure}
     \centering
     \begin{subfigure}[b]{0.2\textwidth}
         \centering
    \includegraphics[scale = 0.25]{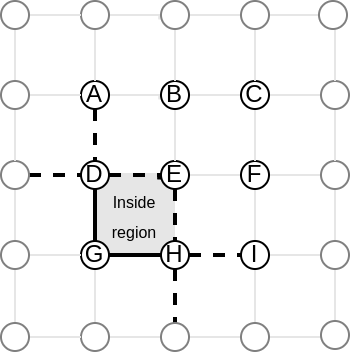}
    \caption{Corner where SBF operation can be applied in a closed region.}
    \label{fig:proof-1}
     \end{subfigure}
     \hfill
     \begin{subfigure}[b]{0.2\textwidth}
         \centering
    \includegraphics[scale = 0.25]{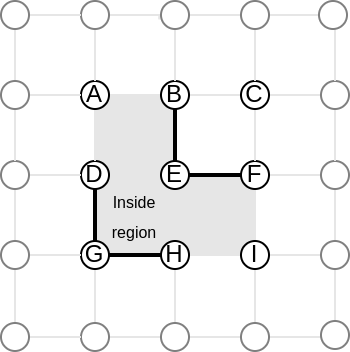}
    \caption{Corner where SBF operation cannot be applied in a closed region }
    \label{fig:proof-2}
     \end{subfigure}
     \hfill
    \caption{Corner in a closed region}
    \label{fig:proof}
\end{figure}

A path between two nodes is \textbf{trivial} if it only moves (either horizontally or vertically via the edges) in the direction of the destination node. Trivial paths are also the shortest paths from the source node to the destination node. 

Non-trivial paths move towards and away from the destination (i.e., reverse their direction) multiple times before reaching the destination node. A path reverses its direction by turning right or left twice consecutively. In such cases, a straight line along the edges can be drawn to connect two nodes on the path such that a closed region is formed. A non-trivial path can be transformed into a trivial path by removing all such closed regions. The procedure to reduce a closed region to a straight line is as follows:

While a closed region exists, perform the following steps.

\begin{enumerate}
    \item Apply SBF operation $S_1$ to all four-node grids inside the closed region. 

    \item Apply SBF operation $S_2$ to all corners in the closed region wherever it is valid.
    
\end{enumerate}
 
In the above procedure, both steps reduce the area inside the closed region. The $S_1$ and $S_2$ operations are shown in the Figure \ref{fig:coverage-2x2}. The above procedure fails to produce an output only when no valid operations can be performed. The following argument shows that there always exists valid operations $S_1$ or $S_2$ till a closed region is present.

In both operations $S_1$ and $S_2$, corners are considered for reducing the area. Let $G$ (refer to Figure \ref{fig:proof} ) be a corner inside the closed region where the SBF operation is performed. The four-node grid under consideration is $DEGH$. The border of the region through $G$ passes through nodes $B$ and $H$ (shown as a solid black line in Figure \ref{fig:proof}), after which it can continue in any direction (dotted black lines in Figure \ref{fig:proof-1}). In either direction, if the border passes through $E$, the operation $S_1$ can be applied.


If the border does not pass through $E$ and if there is no other segment of the border passing through $E$, then the $S_2$ operation operation can be applied to reduce the area. The SBF operation $S_2$ becomes invalid on the four-node grid $DEGH$ if there is another segment of the border passing through $E$. 

If there is another border segment through $E$, it must pass through nodes $B$ and $F$. As the border is continuous and encloses a closed region, $D$ must meet $B$ and $H$ must meet $F$. In both cases, there has to be a corner where they can meet. The $S_2$ operation cannot be invalid because the borders have to meet. Hence, there has to be a corner where the SBF operation is valid. The above procedure ends when no area is left to reduce, i.e., only a straight connecting two nodes is left. 

With the above procedure, any nontrivial path can be transformed into a trivial path. Also, any trivial path can be transformed into any other one by applying multiple $S_2$ operations. As the SBF operations are reversible, any trivial path can be transformed into a non-trivial path. Hence, any valid path can be transformed into any other one, allowing all paths to be explorable by applying SBF operations.

\section{QAOA for Coverage Path Planning}
\label{sec:cpp-qaoa}

In this section, the constructions of QAOA for the multi-vehicle CPP problem are discussed. The techniques of \cite{hadfield2019quantum} are used to design suitable problem encodings and mixing operators such that the probability amplitude is restricted to quantum states encoding feasible solutions. The circuits yield parameterized quantum states that represent possible solutions for multiple robots (via repeated state preparations and measurements). Measuring such a state in the computational basis returns paths for each robot. The methodology for designing initial state, mixing, and phase separation operators is presented in the following sections.

\subsection{Initial state}

The initial state is created by applying basis encoding on all bit strings in the initial population. Thus, the initial state is a superposition of all bit strings in the initial population (refer to Section \ref{sec:initial-population}).

\subsection{Mixers}

Mixers explore feasible solutions for the problem at hand. Hence, mixers that can imitate the mutation process discussed earlier are designed. Specifically, a partial mixer that performs SBF operation and a full mixer for applying SBF operation on all four-node graphs are needed. As the operation must be applied in valid scenarios, controlled-unitary partial mixers\cite{stollenwerk2020toward} are preferable. The design of mixers is discussed in the following sub-section:

\begin{figure}
    \centering
    \scalebox{1}{
        \Qcircuit @C=1em @R=1em {
         &\gate{H} & \ctrl{1} & \qw  & \qw    & \qw   &\qw    &\qw    &\ctrl{1}    &\gate{H}    &\qw    \\
         &\gate{H} & \targ    & \ctrl{1}  & \qw & \qw &\qw    &\ctrl{1}    &\targ    &\gate{H}    &\qw\\
         &\gate{H} & \qw      & \targ     &\ctrl{1}    & \qw & \ctrl{1} &\targ     &\qw    &\gate{H}    &\qw\\
         &\gate{H} & \qw      & \qw     &\targ    & \gate{R_Z(\beta)} & \targ &\qw   &\qw      &\gate{H}    &\qw\\
        }}
    \caption{Simultaneous Bit Flip mixer that flips all the qubits simultaneously and is controlled by a parameter $\beta$.}
    \label{fig:rxxxx-rotation}
\end{figure}
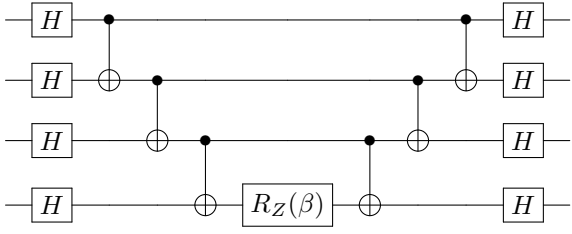

\subsubsection{Simultaneous Bit Flip mixer}
A simultaneous Bit Flip (SBF) mixer performs an SBF operation defined earlier. It has a trainable parameter $\beta$ that controls the probability of the operation being applied. The circuit of the SBF mixer is shown in Figure \ref{fig:rxxxx-rotation}. The SBF mixer is defined as follows:

\begin{equation}
    U_{ab,bc,cd,da}^{SBF}(\beta) = exp(-i \beta (X_{ab} X_{bc} X_{cd} X_{da}) / 2)
\end{equation}

\subsubsection{Partial controlled four-node mixer $U_{a,b,c,d}^{4N}$} A partial mixer $U_{a,b,c,d}^{4N}$ is designed which is the controlled SBF mixer with the logic function $f(a,b,c,d)$ as the controlling logic. This is defined as follows:

\begin{equation}
    U_{a,b,c,d}^{4N}(\beta) =  \wedge_{f(a,b,c,b)} U_{ab,bc,cd,da}^{SBF}(\beta)
\end{equation}

Here, $U_{ab,bc,cd,da}^{SBF}(\beta)$ denotes the SBF mixer, and $f(a,b,c,d)$ is the controlling logic function.

\subsubsection{Full mixer design}
The full mixer for a grid of size $m \times n$ is defined as the product of all partially controlled $U_{a,b,c,d}^{4N}$ Unitaries over all four-node grids present. This is defined as follows:

\begin{equation}
\label{eq:cpp-mixer}
    U_{CPP}(\beta) = \prod_i^{(m-1)(n-1)} U_{a,b,c,d}^{4N}(\beta)
\end{equation}

As it is possible to transform a random feasible path into any other feasible path, the full mixer can cover the entire feasible search subspace. The Boolean condition for each controlled mixer ensures that the new states explored are always inside the feasible subspace. Hence, the full mixer satisfies the conditions required by a mixing operator mentioned in \cite{hadfield2019quantum}.

\subsection{Phase separation operator}
The phase separation operator is encoded as a diagonal cost Hamiltonian $C$ that acts on state $x$ as $C\ket{x} = c(x) \ket{x}$. The problem cost function (see Equation-\ref{eq:cpp-cost}) is mapped to the Pauli Z operator using the transformations described in \cite{hadfield2021representation}. The relevant equations after applying the transformations are shown below.

\begin{multline}
\label{eq:cpp-cost-z1}
    C = \sum_{r,i} w_i (1-Z_{r,i})  + \\ \sum_{r} \left [   \sum_{j} d_j(1-Z_{r,j}) -  \sum_{j} d_j(1-Z_{r+1,j})  \right] ^ 2  + \\ \sum_{r} \sum_j (1-Z_{r,j}) (1-Z_{r+l,j})
\end{multline}

This can be written as Equation \ref{eq:cost-z3} and implemented using $R_Z$ and $R_{ZZ}$ gates.
\begin{equation}
\label{eq:cost-z3}
    C =  c_0 I -
    \sum_{r,i} c_1 Z_{ri} + \sum_{r,i} c_2 Z_{r,i} Z_{r+1,j}  
\end{equation}

\subsection{Resource estimates}

\begin{figure}
    \centering
    \includegraphics[scale=0.2]{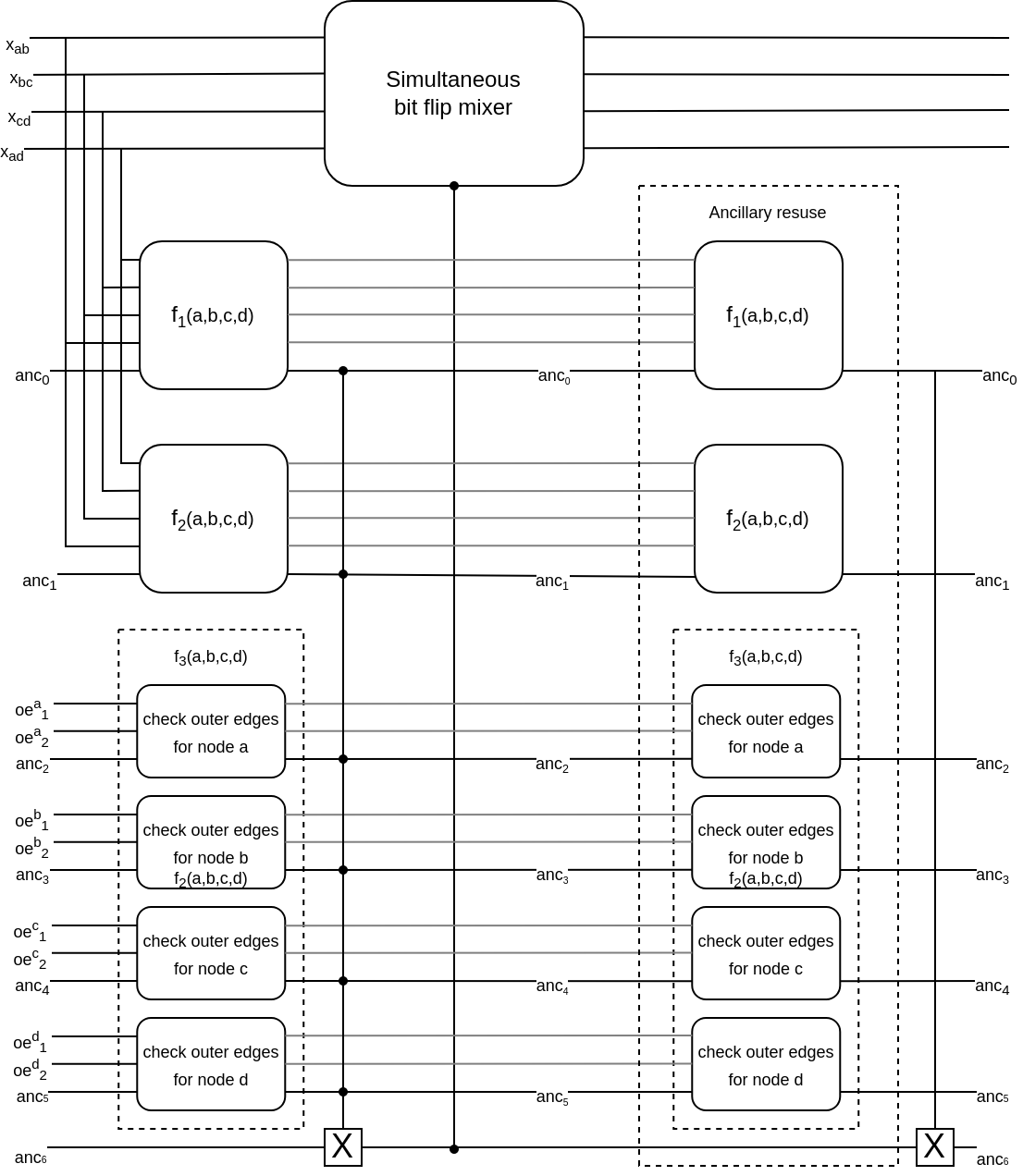}
    \caption{A schematic diagram of a partial QAOA mixer circuit}
    \label{fig:qaoa-schematic-circuit}
\end{figure}

As optimal circuit compilation is hardware-dependent, approximate resource estimates of the proposed QAOA solution are shown here. CNOT and single-qubit gates are considered universal gate sets\cite{wang2020x}. We show an approximate resource estimate in terms of the number of single-qubit gates $\mathbb{N}_S$ and the number of CNOT gates $\mathbb{N}_C$. With simple circuit optimizations, the resource estimates can be further reduced.

The problem formulation requires $r \times e$ qubits where $e = (m (n - 1) + n (m-1))$ is the number of edges in the grid, and $r$ is the number of robots. We also use eight ancillary qubits per robot to compute the logic function controlling the mixer. As we reuse the ancillary qubits for all controlled partial mixers, the number of qubits required is significantly reduced but at the cost of increased circuit depth. Hence, the number of qubits needed is $\mathbb{N}_Q \approx re$.

The resource estimates for the phase separation operator are as follows. Each linear term in Equation \ref{eq:cost-z3} is implemented using an RZ gate requiring a total of $re$ single-qubit gates. Similarly, each quadratic term is implemented using 2 CNOT gates and 1 RZ gate, requiring a total of $re (re + 1)$ CNOT gates and $re (re + 1)/2$ single-qubit gates. 

A Full mixer consists of $r(m-1)(n-1)$ partial mixers. The schematic circuit diagram of a partial mixer is shown in Figure \ref{fig:qaoa-schematic-circuit}. The controlled Z operation used in the partial mixer can be decomposed into 2 Z rotations and 2 CNOT gates \cite{barenco1995elementary}. $f_1(a,b,c,d)$ needs one 5-qubit Tofolli gate, $f_2(a,b,c,d)$ needs two 5-qubit Tofolli gates, and $f_3(a,b,c,d)$ needs four 3-qubit Tofolli gates. A 7-qubit Toffoli gate is used for computing $f(a,b,c,d)$ on the ancilla qubit $anc_6$, which controls the partial mixer (SBF mixer).

An $(l+1)$-qubit Tofolli gate can be decomposed into $4 (l-2)$ 3-qubit Tofolli gates using $l-2$ work qubits, when $l \geq 3$ \cite{barenco1995elementary}. Finally, each 3-qubit Tofolli gate can be decomposed into 6 CNOT and nine single-qubit gates. The ancillary reuse reduces the number of qubits required but requires recomputing all the functions. Hence, the total gate requirement is twice of the controlled partial mixer.

\section{Experiments}
\label{sec:experiments}

The proposed approach was verified by implementing the QAOA and Simulated Annealing algorithm. The results were compared with the Depth First Search (DFS) algorithm. DFS algorithm finds the best solution by performing an exhaustive search on the entire search space. Due to computational constraints, the QAOA algorithm could be tested for a single robot on a 3x3 grid. Hence, only the SA algorithm was tested for multiple robots on larger grids.

\subsection{Collision free path for a single robot on 3x3 grid}

\begin{figure}
     \centering
     \begin{subfigure}[b]{0.2\textwidth}
         \centering
    \includegraphics[scale = 0.2]{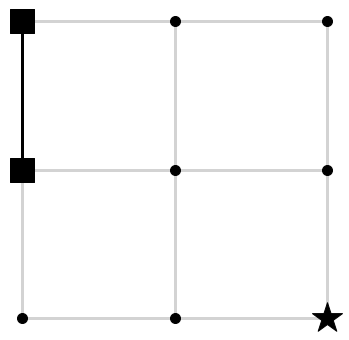}
    \caption{Initial state}
    \label{fig:3x3-init-path}
     \end{subfigure}
     \hfill
     \begin{subfigure}[b]{0.2\textwidth}
         \centering
    \includegraphics[scale = 0.2]{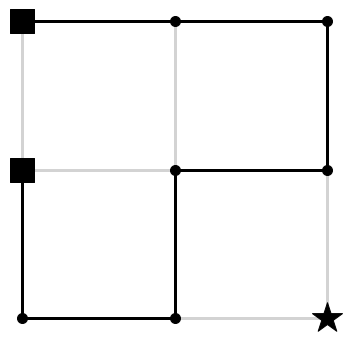}
    \caption{Optimal path}
    \label{fig:3x3-optimal-path}
     \end{subfigure}
     \hfill
    \caption{Initial state and optimal path as computed by QAOA, DFS, and SA}
    \label{fig:3x3-grid-paths}
\end{figure}

In this experiment, it was verified if the QAOA algorithm can find an optimal path for a single robot on a 3x3 grid. The QAOA algorithm was simulated using the Pennylane library. The mixer circuit in the QAOA is independent for each robot. Hence, verifying if the mixer circuit can explore paths for a single robot is sufficient to ensure that it can explore paths for multiple robots.

Twenty qubits (12 decision variable qubits + 8 ancillary qubits) were needed to simulate the QAOA algorithm for this case. The parameters were optimized using gradient descent with the Nesterov momentum algorithm (keeping step size = 0.1). The initial state and the optimal path returned by the QAOA algorithm are shown in Figure~\ref{fig:3x3-init-path} and Figure~\ref{fig:3x3-optimal-path}. The square nodes represent the endpoints, while the triangular nodes represent the obstacles.

\begin{figure}
     \centering
     \begin{subfigure}[b]{0.22\textwidth}
        \centering
        \includegraphics[scale = 0.28]{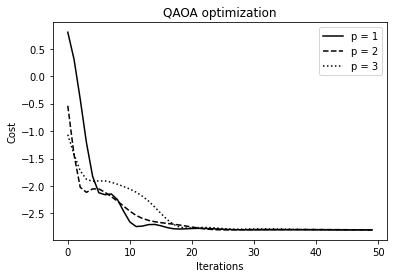}
        \caption{QAOA cost plot.}
        \label{fig:qaoa-loss-plot}
     \end{subfigure}
     \hfill
     \begin{subfigure}[b]{0.22\textwidth}
        \centering
        \includegraphics[scale = 0.24]{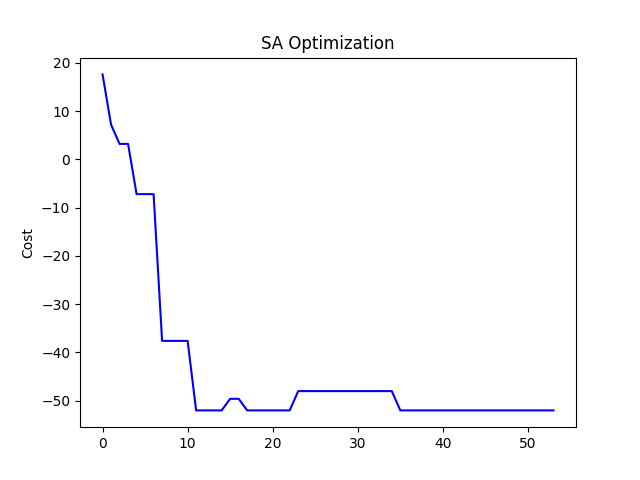}
        \caption{SA cost plot.}
        \label{fig:sa-loss-plot}  
     \end{subfigure}
     \hfill
    \caption{Optimization plots}
    \label{fig:optimization plots}
\end{figure}



The quality of the solution returned by QAOA can be improved by increasing the number of layers $p$. But this also increases the circuit depth and the number of training parameters. The loss curve for different numbers of layers $p$ is plotted in Figure \ref{fig:qaoa-loss-plot}. Due to the simplicity of the scenario, the optimal solution is found with just a single layer. The result is verified to be the same as the result returned by DFS and SA algorithm.


\subsection{Collision-free path for two robots}

\begin{figure}
     \centering
     \begin{subfigure}[b]{0.22\textwidth}
         \centering
         \includegraphics[scale=0.25]{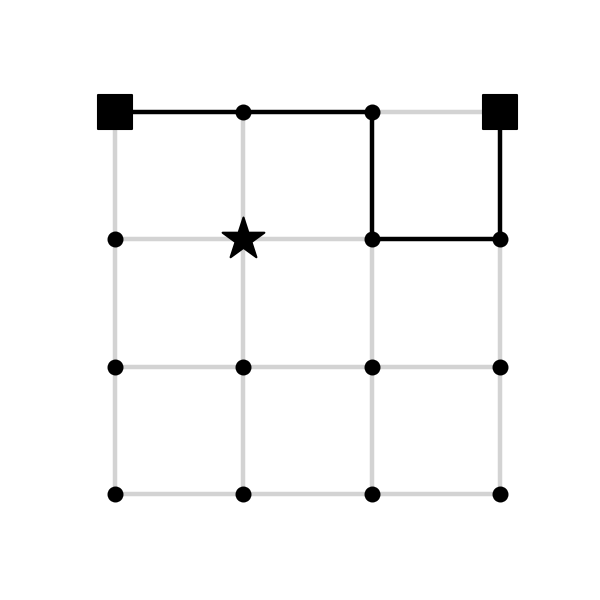}
         \caption{Optimal path for robot 1}
         \label{fig:4x4-robot-1}
     \end{subfigure}
     \hfill
     \begin{subfigure}[b]{0.22\textwidth}
         \centering
         \includegraphics[scale = 0.25]{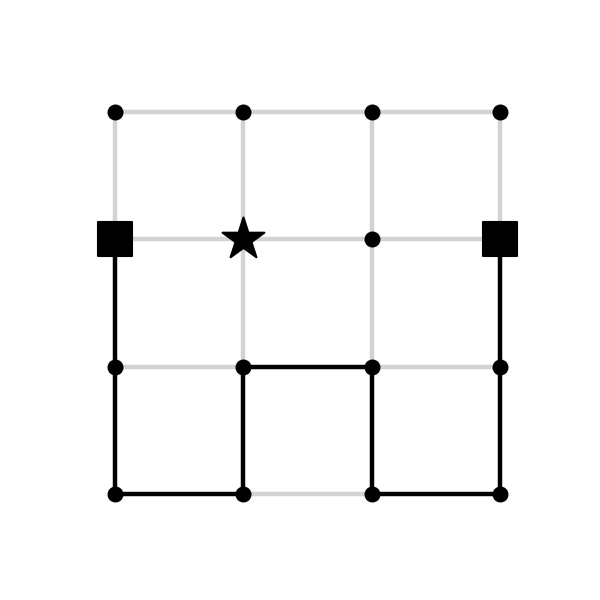}
         \caption{Optimal path for robot 2}
         \label{fig:4x4-robot-2}
     \end{subfigure}
     \hfill
    \caption{Optimal paths for two robots on 4x4 grid}
    \label{fig:4x4-grid-paths}
\end{figure}

\begin{figure}
     \centering
     \begin{subfigure}[b]{0.2\textwidth}
         \centering
         \includegraphics[scale = 0.25]{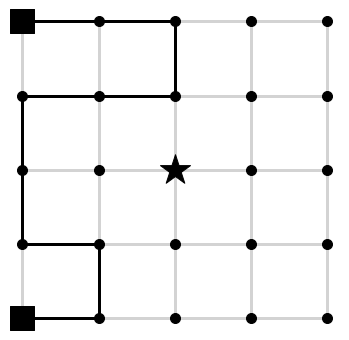}
         \caption{Optimal path for robot 1}
         \label{fig:5x5-robot-1}
     \end{subfigure}
     \hfill
     \begin{subfigure}[b]{0.2\textwidth}
         \centering
         \includegraphics[scale = 0.25]{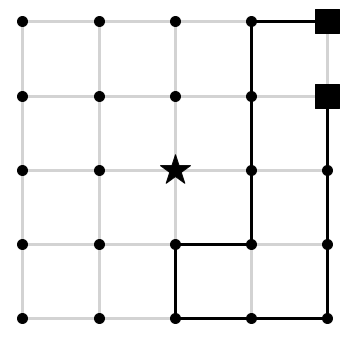}
         \caption{Optimal path for robot 2}
         \label{fig:5x5-robot-2}
     \end{subfigure}
     \hfill
    \caption{Optimal paths for two robots on 5x5 grid}
    \label{fig:5x5-grid-paths}
\end{figure}

The number of qubits needed to run QAOA algorithm for two robots on a 3x3 grid ($\sim\,$40) is already high enough to be simulated classically. Hence, only the SA algorithm was tested and compared with the DFS algorithm for multiple robots. The solutions evaluated by SA for an example scenario of two robots in a 4x4 grid and 5x5 grids are shown in Figure~\ref{fig:4x4-grid-paths} and Figure~\ref{fig:5x5-grid-paths}. The SA loss curve for the 4x4 scenario is shown in Figure \ref{fig:sa-loss-plot}. These are the same results returned by the DFS algorithm, thus verifying the results.



\section{Discussion and conclusion}
\label{sec:discussion}

A novel method for exploring paths in a 2D grid was presented in this work. The methodology was designed such that it could be adapted to quantum heuristic algorithms such as QAOA and classical algorithms such as Evolutionary Algorithms. The circuit constructions for the QAOA algorithm were presented along with approximate resource estimates. The proposed methodology was evaluated by implementing QAOA and SA algorithms and compared with the DFS algorithm.

This work may be extended by formulating this problem as a Quadratic Unconstrained Binary Optimization (QUBO) problem and comparing the results with annealer-based solutions.

\bibliography{ref}

\begin{thebibliography}{10}

\bibitem{almadhoun2019survey}
R.~Almadhoun, T.~Taha, L.~Seneviratne, and Y.~Zweiri, ``A survey on multi-robot coverage path planning for model reconstruction and mapping,'' {\em SN Applied Sciences}, vol.~1, no.~8, pp.~1--24, 2019.

\bibitem{tan2021comprehensive}
C.~S. Tan, R.~Mohd-Mokhtar, and M.~R. Arshad, ``A comprehensive review of coverage path planning in robotics using classical and heuristic algorithms,'' {\em IEEE Access}, 2021.

\bibitem{sun2019multi}
R.~Sun, C.~Tang, J.~Zheng, Y.~Zhou, and S.~Yu, ``Multi-robot path planning for complete coverage with genetic algorithms,'' in {\em International Conference on Intelligent Robotics and Applications}, pp.~349--361, Springer, 2019.

\bibitem{senthilkumar2008spanning}
K.~Senthilkumar and K.~Bharadwaj, ``Spanning tree based terrain coverage by multi robots in unknown environments,'' in {\em 2008 Annual IEEE India Conference}, vol.~1, pp.~120--125, IEEE, 2008.

\bibitem{agmon2006constructing}
N.~Agmon, N.~Hazon, and G.~A. Kaminka, ``Constructing spanning trees for efficient multi-robot coverage,'' in {\em Proceedings 2006 IEEE International Conference on Robotics and Automation, 2006. ICRA 2006.}, pp.~1698--1703, IEEE, 2006.

\bibitem{gao2019stc}
G.-Q. Gao and B.~Xin, ``A-stc: auction-based spanning tree coverage algorithm formotion planning of cooperative robots,'' {\em Frontiers of Information Technology \& Electronic Engineering}, vol.~20, no.~1, pp.~18--31, 2019.

\bibitem{huang2018potential}
C.~Huang, W.~Li, C.~Xiao, B.~Liang, and S.~Han, ``Potential field method for persistent surveillance of multiple unmanned aerial vehicle sensors,'' {\em International journal of distributed sensor networks}, vol.~14, no.~1, p.~1550147718755069, 2018.

\bibitem{rosa2020using}
R.~Rosa, T.~Brito, A.~I. Pereira, J.~Lima, and M.~A. Wehrmeister, ``Using multi-uav for rescue environment mapping: Task planning optimization approach,'' in {\em Portuguese Conference on Automatic Control}, pp.~507--517, Springer, 2020.

\bibitem{zhang2019optimal}
X.~Zhang, S.~Liu, and Z.~Xiang, ``Optimal inspection path planning of substation robot in the complex substation environment,'' in {\em 2019 Chinese Automation Congress (CAC)}, pp.~5064--5068, IEEE, 2019.

\bibitem{huang2018development}
Y.-C. Huang and H.-Y. Lin, ``Development and implementation of a multi-robot system for collaborative exploration and complete coverage,'' in {\em 2018 14th International Conference on Signal-Image Technology \& Internet-Based Systems (SITIS)}, pp.~472--479, IEEE, 2018.

\bibitem{nair2020gm}
V.~G. Nair and K.~Guruprasad, ``Gm-vpc: An algorithm for multi-robot coverage of known spaces using generalized voronoi partition,'' {\em Robotica}, vol.~38, no.~5, pp.~845--860, 2020.

\bibitem{caliskanelli2014multi}
I.~Caliskanelli, B.~Broecker, and K.~Tuyls, ``Multi-robot coverage: A bee pheromone signalling approach,'' in {\em Artificial Life and Intelligent Agents Symposium}, pp.~124--140, Springer, 2014.

\bibitem{de2015bio}
F.~De~Rango, N.~Palmieri, X.~S. Yang, and S.~Marano, ``Bio-inspired exploring and recruiting tasks in a team of distributed robots over mined regions,'' in {\em 2015 International Symposium on Performance Evaluation of Computer and Telecommunication Systems (SPECTS)}, pp.~1--8, IEEE, 2015.

\bibitem{palmieri2015multi}
N.~Palmieri, F.~De~Rango, X.~S. Yang, and S.~Marano, ``Multi-robot cooperative tasks using combined nature-inspired techniques,'' in {\em 2015 7th International Joint Conference on Computational Intelligence (IJCCI)}, vol.~1, pp.~74--82, IEEE, 2015.

\bibitem{albina2019hybrid}
K.~Albina and S.~G. Lee, ``Hybrid stochastic exploration using grey wolf optimizer and coordinated multi-robot exploration algorithms,'' {\em IEEE Access}, vol.~7, pp.~14246--14255, 2019.

\bibitem{palmieri2019comparison}
N.~Palmieri, X.-S. Yang, F.~De~Rango, and S.~Marano, ``Comparison of bio-inspired algorithms applied to the coordination of mobile robots considering the energy consumption,'' {\em Neural Computing and Applications}, vol.~31, no.~1, pp.~263--286, 2019.

\bibitem{stollenwerk2020toward}
T.~Stollenwerk, S.~Hadfield, and Z.~Wang, ``Toward quantum gate-model heuristics for real-world planning problems,'' {\em IEEE Transactions on Quantum Engineering}, vol.~1, pp.~1--16, 2020.

\bibitem{hadfield2019quantum}
S.~Hadfield, Z.~Wang, B.~O’Gorman, E.~G. Rieffel, D.~Venturelli, and R.~Biswas, ``From the quantum approximate optimization algorithm to a quantum alternating operator ansatz,'' {\em Algorithms}, vol.~12, no.~2, p.~34, 2019.

\bibitem{farhi2014quantum}
E.~Farhi, J.~Goldstone, and S.~Gutmann, ``A quantum approximate optimization algorithm,'' {\em arXiv preprint arXiv:1411.4028}, 2014.

\bibitem{hadfield2021representation}
S.~Hadfield, ``On the representation of boolean and real functions as hamiltonians for quantum computing,'' {\em ACM Transactions on Quantum Computing}, vol.~2, no.~4, pp.~1--21, 2021.

\bibitem{wang2020x}
Z.~Wang, N.~C. Rubin, J.~M. Dominy, and E.~G. Rieffel, ``X y mixers: Analytical and numerical results for the quantum alternating operator ansatz,'' {\em Physical Review A}, vol.~101, no.~1, p.~012320, 2020.

\bibitem{barenco1995elementary}
A.~Barenco, C.~H. Bennett, R.~Cleve, D.~P. DiVincenzo, N.~Margolus, P.~Shor, T.~Sleator, J.~A. Smolin, and H.~Weinfurter, ``Elementary gates for quantum computation,'' {\em Physical review A}, vol.~52, no.~5, p.~3457, 1995.

\end{thebibliography}
\bibliographystyle{ieeetr}

\end{document}